\documentclass[conference]{IEEEtran}
\IEEEoverridecommandlockouts
\usepackage{cite}
\usepackage{amsmath,amssymb,amsfonts}
\usepackage{algorithmic}
\usepackage{graphicx, subcaption}
\usepackage{subfig}
\usepackage{textcomp}
\usepackage{xcolor}
\usepackage{verbatim}
\usepackage{graphicx}
\usepackage{cite}
\usepackage{booktabs}
\usepackage{placeins}
\usepackage{tikz}
\usepackage{amssymb}
\usepackage{pifont}
\usepackage{tabularx}
\usepackage{multirow}
\usepackage[]{mdframed}
\usepackage{mathtools}
\usepackage{makecell}
\newcommand{\xmark}{\text{\ding{55}}}
\def\BibTeX{{\rm B\kern-.05em{\sc i\kern-.025em b}\kern-.08em
    T\kern-.1667em\lower.7ex\hbox{E}\kern-.125emX}}

\begin{document}

\title{Conference Paper Title*\\
{\footnotesize \textsuperscript{*}Note: Sub-titles are not captured in Xplore and
should not be used}
\thanks{Identify applicable funding agency here. If none, delete this.}
}

\title{Priceless: An examination of Serverless Functions-as-a-Service (FaaS) pricing models}

\author{\IEEEauthorblockN{Nnamdi Ekwe-Ekwe}
\IEEEauthorblockA{
\textit{Ive Sent It LLC}\\
30 N Gould Street, STE R, \\ Sheridan, WY 82801 \\
nnamdi.ekwe-ekwe@ivesentit.org}


}

\maketitle

\begin{abstract}
Serverless Functions-as-a-Service providers have grown in their offering since inception a decade ago, with a myriad of new functionalities offered to end-users. These new features have also brought new, varied and at times complex pricing models that differ between providers. In this paper, we present \textit{Priceless}, a detailed examination of the current state of the art of FaaS features, and their pricing models. We then perform a comparative price analysis of running example workloads across AWS Lambda, Microsoft Azure Functions and Google Cloud Functions. Our work finds significant cost differences both cross-provider, but also, cross-region within the provider. We find that AWS is the cheapest overall to run functions on, with Microsoft Azure being the most expensive for equivalent workloads.
\end{abstract}


\section{Introduction} \label{introduction}

Serverless computing is a cloud computing paradigm, where users deploy their code/applications to a service provider without having to manage any of the underlying infrastructure \cite{shafiei2022serverless, wen2023rise, li2022serverless}. Functions-as-a-Service (FaaS) is arguably the most well-known segment of the serverless paradigm, in which users develop their application as a function (or group of functions) in a particular language, on a specific FaaS cloud provider, with their code packaged and deployed for end-user use \cite{ekwe2024state}. The end-user does not need to be concerned about managing any of the underlying architecture, with their application code run or invoked in response to requests \cite{ekwe2024state,li2022serverless, van2018serverless}. A still maturing field, the first public serverless provider was introduced in 2015 (AWS Lambda\cite{awslambda}). Over a decade later, this has grown significantly, with numerous public serverless offerings available from most of the top cloud computing companies \cite{ekwe2024state}. This paradigm has brought with it its own unique pricing models which differ from the traditional cloud \cite{cloudflareserverless, liu2023demystifying}. Whilst in traditional cloud Infrastructure-as-a-Service (IaaS) pricing models, users pay for a continuously running instance \cite{wu2019cloud}, FaaS has a different paradigm, with users paying only when their function is running \cite{adzic2017serverless}. FaaS pricing is typically based on two main factors - (1) the price of using the resources allocated to the function (whether that be CPU and/or RAM) in addition to (2) an invocation cost for the function \cite{ekwe2024state}.\\ 

Functions-as-a-Service is still a growing and maturing field \cite{datadogserverlessreport, ekwe2024state} and the functionality offered by the FaaS providers has been significantly expanded since initial inception. Providers have extended their feature set to include options such as pre-provisioned functions \cite{azurefunctionsflexconsumptionplan, awsprovisionedconcurrency} to deal with the cold start problem \cite{manner2018cold, chai2025fork}, instance-backed FaaS \cite{aws_lambda_managed_instances, aws_lambda_managed_instances_blog, azure_functions_dedicated_plan} to provide greater resources for function execution or even Durable Functions to enable more complex FaaS workflows \cite{azure_durable_functions_overview, aws_lambda_durable_functions} to name but a few. This significant expansion across providers' offerings have led to new pricing models being created to charge for the usage of such features \cite{microsoftazurefunctionspricing, awslambdapricing, googlecloudfunctionspricing}. The sub-field of pricing in FaaS is maturing and as such still limited in scope, with many open research areas and challenges still to resolve \cite{ghasemi2025dynamic}. As of yet however, there has been limited work capturing (1) the current state of the art of all the new features offered by these FaaS providers, in addition to (2) performing a detailed qualitative and quantitative exploration of their accompanying pricing models, and their potential effects on function workload cost. Whilst in the industry-focused world, there have been articles written by developers examining pricing and comparing costs across providers \cite{raymunene_lambda_provisioned_concurrency, wiz_azure_vs_aws_cloud_cost, synextra_aws_vs_azure_functions_2025}, such work is limited in scope and disparate in nature with a cohesive, detailed examination/comparison of FaaS pricing models still required.\\

In this paper therefore, we introduce \textit{Priceless}, an analysis of three well known FaaS providers (AWS Lambda \cite{awslambda}, Google Cloud Run Functions \cite{googlecloudfunctions} and Microsoft Azure Functions \cite{microsoftazurefunctions}) exploring all their features, and their accompanying pricing models based on the providers' publicly available information. Our work aims to (1) present the state of the art of all the current features offered by these providers, (2) present and model their accompanying pricing models and then (3) perform an empirical analysis and comparison of the estimated cost of running workloads cross provider with the intent to inform the end-user of estimated function execution cost differences. Our paper in a new addition to the state of the art, also explores cross regional pricing differences employed by the providers and the accompanying impact that also has on workload cost. Our paper is structured as follows - in Section \ref{faas-feature-set}, we first detail the FaaS state of the art feature set offered by the various providers. Next, in Section \ref{pricing-models}, we explore and model the pricing strategies employed by the providers for their features. In Section \ref{analysis}, we then perform a quantitative pricing analysis of the estimated cost of different workloads running across the various providers, charting cost differences per provider as well as per regions between providers. In Section \ref{observations}, we make a number of observations from our study before discussing related work in Section \ref{relatedwork} and then concluding in Section \ref{conclusion}.

\section{The Current FaaS Feature Set} \label{faas-feature-set}

AWS is the most mature of all the platforms having been the first to launch in 2015 \cite{lambdalaunch}. This was followed by Azure functions in 2016 \cite{azurefunctionsga} with Google releasing their own serverless offering in 2018 \cite{gcffunctionsga}. This timeline has translated into the breadth of the features offered by the platforms with AWS Lambda being the most feature-rich of all the FaaS providers \cite{awslambda}. The core functionalities of users being able to write their function in a particular language, allocate resources, select an invocation type (such as HTTP, queue-based triggers, etc.) allocate a timeout for function execution and then process requests are all supported by the providers (albeit with some differences around the breadth of functionality) \cite{ekwe2024state}. In recent years however, the platforms have sought to add more advanced features to their core offerings to allow for different kinds of applications as well as solve inherent limitations with FaaS as a paradigm. We detail these next.

\subsection{Pre-Provisioned Functions} \label{preprovisionedfunctions}
With FaaS, execution environments are dynamically created to handle incoming function requests and are subsequently terminated once requests complete \cite{shafiei2022serverless}. This is economically advantageous for the consumer, but introduces a unique issue to FaaS known as the ``cold-start'' problem \cite{manner2018cold}. When a request is received, an execution environment/container is spawned on the underlying hardware to process that request. This environment isn't immediately available and as such, the time it takes to create that environment on-demand to process the end-user request is known as a ``cold-start'' \cite{manner2018cold, chai2025fork}. Once a request completes, the environment is not immediately terminated but remains available or ``warm'' to be able to process subsequent end-user requests. After a period however, the execution environment is terminated with the same ``cold start'' issue occurring again for future requests \cite{manner2018cold}. Mitigating cold-starts has created its own body of research with many works exploring how to effectively deal with this issue \cite{ristov2022colder, liu2023faaslight, shen2021defuse, silva2020prebaking}. FaaS providers have also introduced features to resolve this problem with their services.\\

Microsoft introduced a feature allowing users to deploy functions in a ``dual-mode'' setting, where users can use the standard on-demand functionality as normal \cite{azurefunctionsflexconsumptionplan}, but additionally utilise ``always-ready'' instances to reduce cold-start delay. These ``always-ready'' instances are an optional extra for the user and mean that they can select a specific number of instances to always be provisioned and ready to process requests from end-users \cite{azurefunctionsflexconsumptionplan}. If additional resources are needed, then standard on-demand instances can be used. Microsoft even offers a further plan - the Premium Plan, in which users don't have to use the on-demand serverless at all, instead paying for always-ready instances permanently \cite{azurefunctionspremiumplan}. Google has a similar feature providing for minimum instances to process requests which can then utilize standard on-demand instances when more resources are required \cite{googlecloudrun_min_instances}. AWS's equivalent feature is called Provisioned Concurrency \cite{awsprovisionedconcurrency}, a feature allowing for FaaS environments to be pre-provisioned and ``always-on''. This means that function invocations would not have an initial cold start but rather be processed by the pre-provisioned online instances \cite{awsprovisionedconcurrency}. With Provisioned Concurrency, the user continuously pays for the number of provisioned concurrency instances they deploy. If function invocations exceed the configured resources assigned to those provisioned environments, then standard on-demand instances are used to process additional requests. 

\subsection{The expansion of FaaS with traditional IaaS} \label{faasiaas}

The FaaS paradigm has led to several proposed application domains that could take advantage of the paradigm \cite{shafiei2022serverless}. There are however certain application domains inherently unsuited to FaaS due to its relatively constrained resources, inherent limits on function execution time, or lack of additional hardware available \cite{shafiei2022serverless}. In recent years, the FaaS providers have ameliorated their offerings by bringing the power of the traditional IaaS to serverless. AWS in October of 2025 introduced Lambda Managed Instances \cite{aws_lambda_managed_instances_blog, awslambdamanagedinstances} which uses the flexibility of Lambda and FaaS backed by the power of traditional IaaS machines. This feature allows users to run workloads on dedicated infrastructure with the flexibility of FaaS whilst having access to features such as high-bandwidth or significantly more powerful compute options \cite{aws_lambda_managed_instances_blog}. Microsoft Azure also have similar offerings with Dedicated Plans and Container Apps that allow users to run functions on dedicated VMs \cite{azurescaleandhosting}. Google however does not have an equivalent of this feature with users having to deploy their functions on the standard FaaS infrastructure. 

\subsection{Accommodating complex scenarios} \label{complexscenarios}

FaaS is inherently stateless, posing a challenge when having to build stateful or fault-tolerant workloads \cite{burckhardt2021durable}. Additionally, coordinating or building fully fledged applications comprising multiple functions is also a challenge on the current FaaS paradigm.  Providers such as AWS Lambda introduced Lambda Durable Functions \cite{aws_lambda_durable_functions_doc}, allowing users to build multi-step applications (comprising multiple different individual functions) that are resilient, enabling checkpointing and automatic replay/retry from failure. Durable Functions make use of progress tracking, storage and checkpoints that can be used in domains such as payments, human-in-the-loop function applications, complex multi-step order fulfillment, or other complex workflows \cite{aws_lambda_durable_functions_doc}. AWS makes an API available to the end user exposing simple primitives that the user can invoke such as \textit{waits}, \textit{retries}, etc. into their applications using standard programming languages like Python or Typescript \cite{aws_lambda_durable_functions_doc}. These primitives allow function execution to replay past executions, ignoring completed steps by using stored results rather than re-executing past operations, etc. Azure have a comparable offering in their own Azure Durable Functions \cite{azure_durable_functions_overview} that offers similar functionalities allowing for features such as function chaining, human-in-the-loop functions or dealing with external asynchronous dependencies, to name but a few. Azure provides four different types of functions that act as building blocks for the user to deploy their applications on the Durable paradigm \cite{azure_durable_functions_types_features}. Similar to the IaaS-backed instances, Google does not have a Durable Functions equivalent on their FaaS offering.

\subsection{Additional features} \label{additionalfeatures}

AWS (being the most mature/feature-rich of the three providers) offers a significant set of additional features that add to their overall end-user FaaS deployments \cite{awslambdapricing}. Whilst these functionalities are only specific to AWS with equivalents not available in the other three platforms, these are important features to note.\\

AWS offers the use of both x86 and ARM (Graviton backed) Lambda environments with customers able to deploy FaaS functions on any of the two \cite{aws_graviton2_lambda}. AWS states that deploying on ARM backed Lambdas also has the added advantage of generally being cheaper and more performant for workloads \cite{aws_lambda_arm_vs_x86}. AWS offers the use of Tenant Isolation \cite{aws_lambda_tenant_isolation} enabling strict environment isolation for separate tenants. This can be advantageous in scenarios where for example, an application developer wants to ensure that tenant specific data is processed in a specific environment to that particular tenant. AWS added Lambda Ephemeral Storage \cite{aws_lambda_ephemeral_storage} to their offering, allowing the developer to make use of additional storage (at additional cost) for their functions. Such extra storage allows for use cases like data storage for processing jobs (such as files downloaded from S3), ETL jobs or even Machine Learning workloads. As another means of addressing latency issues in functions, AWS introduced SnapStart \cite{aws_lambda_snapstart}, also designed to address latency in executing functions by performing a snapshot of a function's memory and disk state and caching it for faster access \cite{awslambdapricing}. SnapStart differs from Provisioned Concurrency discussed earlier, as it targets faster loading of modules or frameworks a function depends on, reducing the latency from several seconds to sub-second latency \cite{aws_lambda_snapstart}. This feature, AWS states is more advantageous for functions invoked frequently and at scale. Finally, AWS have other features available such as handling Lambda functions that return an HTTP response stream, Lambdas deployed at the edge (using Amazon CloudFront) or even handling traffic spikes in specific AWS Lambda functions that are triggered by queues \cite{awslambdapricing}, all at extra cost.\\

\begin{table}
\caption{FaaS feature offering}
\begin{center}
\begin{tabular}{|c|c|c|c|}
\hline
\textbf{Feature} & \textbf{AWS} & \textbf{Microsoft} & \textbf{Google} \\
\midrule
\hline
Standard FaaS functionality & \checkmark & \checkmark & \checkmark \\\hline
Pre-Provisioned FaaS instances & \checkmark & \checkmark & \checkmark \\\hline
IaaS backed FaaS & \checkmark & \checkmark & \xmark \\\hline
Durable Functions & \checkmark & \checkmark & \xmark \\\hline
\end{tabular}
\end{center}
\label{tab:faasfeaturecomparison}
\end{table}

We have detailed the full feature set of the three platforms and have identified common features supported. Next, we model and discuss the pricing strategies employed by the providers.

\section{Pricing Models} \label{pricing-models}

\subsection{Founding Principles}

The underpinning pricing principle in FaaS is to charge for the duration of use of the resources allocated to the function, in addition to a flat invocation fee \cite{awslambdapricing, googlecloudfunctionspricing, microsoftazurefunctionspricing, ekwe2024state}. The FaaS providers give the resources a user can allocate to their function in MB or MiB, with execution duration expressed in milliseconds. The user chooses appropriate resources to allocate (either the CPU/memory or both) with their function consuming said resources during execution. The cost comprises the following:\\

\begin{enumerate}
    \item \textbf{Compute usage}: The unit of duration in FaaS is known as a \textit{GB-second} (or \textit{GiB second} depending on provider). Resources allocated to a function are typically displayed by the providers in MB, and so the gigabyte value of the resources - \(r\) allocated to a function is obtained by converting the initial MB value to GB. This GB value \(r\) is then multiplied by the duration \(d\) (given by the provider in milliseconds) of the request but then converted to seconds. This is finally multiplied by the price point \(p_{mem}\) per GB-second for the memory.
    \item \textbf{An invocation fee}: This is additionally charged by the provider, which is based the price per request invocation \(p_{inv}\).\\
\end{enumerate}

We can therefore express the total cost function \(C_{DEF}\) of running a FaaS function as follows (Equation \ref{eqn:faaspricing}):

\begin{equation}
\label{eqn:faaspricing}
\begin{aligned}
t(d) 
&= \frac{d}{1000} \\[1ex]
C_{DEF}(r, d)
&=
\underbrace{\frac{r}{1024} \cdot t(d) \cdot p_{mem}}_{\text{Compute}}
+
\underbrace{p_{inv}}_{\text{Invocation}}
\end{aligned}
\end{equation}

where \(C_{DEF}\) is the cost function that takes the size of resources requested \(r\) and the duration of the request \(d\). AWS Lambda \cite{awslambdapricing} and Microsoft Azure's Consumption Plan \cite{azurefunctionsconsumptionplan} follow Equation \ref{eqn:faaspricing} for standard, on-demand use of their FaaS platforms. Note that we divide \(r\) by 1024 to get the \textit{gigabyte} representation. We also use the function \(t(d)\) to calculate the billed duration time as some providers calculate the billed time differently. For AWS and Microsoft Azure's Consumption Plan, billing is done to the nearest ms and so in this case \(t(d)\) is divided by 1000 to obtain the billed duration time (in seconds). For both platforms the resource \(r\) is for memory only, allowing the user to modify only the amount of memory they can assign a function. In the background, CPU is assigned by the FaaS provider transparently, with CPU cores assigned proportionally based on memory selected \cite{awslambdaconfig}.\\

Although Microsoft has their standard Consumption Plan, as of time of writing this plan is set to be replaced by the newer and currently available Microsoft Flex Consumption Plan by 2028 \cite{azurefunctionsconsumptionplan}. This newer Flex Consumption Plan comes with new features, but also changes its pricing model. Whilst it keeps most of Equation \ref{eqn:faaspricing}, the duration calculation \(t(d)\) is no longer to the nearest millisecond. Microsoft charges a \textbf{minimum} execution duration of up to 1s (regardless of actual function execution time) and then, subsequently, charges in 100ms increments for anything over this time. We can therefore express the updated duration, \(t(d)\) and cost equation \(C_{{FLEX}}(r, d)\) as follows:\\

\begin{equation}
\label{eqn:microsoftflex}
\begin{aligned}
t(d) &=
\begin{cases}
1 & \text{if } d \leq 1000 \\
\left\lceil \frac{d}{100} \right\rceil \cdot 0.1 & \text{if } d > 1000
\end{cases} \\[2mm]
C_{\text{FLEX}}(r, d)
&=
\underbrace{\frac{r}{1024} \cdot t(d) \cdot p_{mem}}_{\text{Compute}}
+ \underbrace{p_{inv}}_{\text{Invocation}}
\end{aligned}
\end{equation}

Google Cloud Run Functions (GCF) takes a conceptually similar, yet slightly different approach. First, Google is the only provider of our list to give its resources in MiB (mebibyte) and calculations are in GiB-seconds (gibibyte). Next, Google Cloud has two billing models \textit{request-based} and \textit{instance-based} billing \cite{googlecloudrun_billing_settings}. For \textit{request-based} billing, the user is charged only when the function is executing/processing requests similar to traditional FaaS pricing. With \textit{instance-based} billing however, the user is charged for the entire usage of the instance (even when requests are not being processed). The recommendation from Google is that \textit{instance-based} billing is better for more steady, predictable workloads whilst \textit{request-based} is for unpredictable, ``bursty'' style workloads \cite{googlecloudrun_billing_settings}. Google, similar to Azure, charges to the nearest 100ms \cite{googlecloudfunctionspricing} meaning that the duration element of our equation \(t(d)\) is adapted to become:

\begin{equation}
    t(d) = \left\lceil \frac{d}{100} \right\rceil \cdot 0.1
\end{equation}

Google also allows users to change \textit{both} the CPU and Memory allocations with both of them attracting different prices. If the user configures \textit{request-based} billing, then an invocation fee is charged, however with \textit{instance-based} billing this isn't included. We therefore modify our Equation \ref{eqn:faaspricing} and expand our notations to capture the new elements of the cost function for Google Cloud Functions (GCF):

\begin{equation}
\label{eqn:faaspricing_google_combined}
\begin{aligned}
C_{GCF}^{RB}(r,d)
&=
t(d) \cdot
\underbrace{\Bigg(
      \frac{r}{1024} \cdot p_{mem}
      + c \cdot p_{cpu}
\Bigg)}_{\text{Memory + CPU}}
\\[1ex]
+ \underbrace{p_{inv}}_{\text{Invocation}}
\\[2ex]
C_{GCF}^{IB}(r,d)
&=
t(d) \cdot
\underbrace{\Bigg(
      \frac{r}{1024} \cdot p_{mem}
      + c \cdot p_{cpu}
\Bigg)}_{\text{Memory + CPU}}
\end{aligned}
\end{equation}

where \(p_{mem}\) is the price of the memory, \(p_{cpu}\) is the price of the CPU core, \(c\) denotes the number of CPU cores allocated to the function, and \(IB\) and \(RB\) denote instance-billing and request-billing respectively.\\

Similar to Microsoft, Google also has an older billing model, called the Google Cloud Run Functions (1st gen) pricing model \cite{google_cloudrun_functions_pricing_1stgen}. Google Cloud Run Functions (1st gen) has a similar pricing model to the newer Google Cloud Run Functions and follows Equation \ref{eqn:faaspricing_google_combined}. There are key differences however amongst the two models. First, instance-based billing is not supported by Google cloud Run Functions (1st gen). Also, the units of measurement when it comes to cost calculation are different, with GB-seconds and GHz seconds used to denote memory and CPU calculation whilst this isn't the case with the newer Google Cloud Run Functions (GiB-seconds and vCPU seconds respectively).\\

These three Equations (\ref{eqn:faaspricing}), (\ref{eqn:faaspricing_google_combined}) and (\ref{eqn:microsoftflex}) represent the pricing formulae for on-demand use of FaaS for Azure, AWS and Google Functions. As we discussed in Section \ref{faas-feature-set}, the core features offered by these providers have expanded, leading to new pricing models being created to charge for these new services. We explore these new features and their associated pricing next.

\subsection{Pre-Provisioned Functions} \label{pre-provisioned-functions-models}

\subsubsection{Microsoft}

Using the old Microsoft Consumption Plan, cold starts can still occur as the execution environments (instances) can scale to zero meaning that future requests could be delayed by having to provision a new instance to handle an end-user request \cite{azurefunctionsplans}. Microsoft introduced the Flex Consumption Plan \cite{azurefunctionsflexconsumptionplan} which can use a ``dual-mode'' setting, allowing users to pre-provision a number of ``always-ready'' instances to reduce cold-start delay \cite{azurefunctionsflexconsumptionplan}. Users have the choice of using the Flex Consumption Plan to solely utilise standard on-demand instances as we mentioned in Equation \ref{eqn:microsoftflex}, or additionally enable the ``always-ready'' mode to make use of the pre-provisioned instances. If additional resources are needed (if ``always-ready'' is enabled), then standard on-demand instances are used. For the ``always-ready'' mode, Microsoft continuously bills for the usage of the provisioned resources known as the ``baseline'' \cite{azurefunctionsflexconsumptionplan}. Note that this is billed regardless of \textit{actual} usage (charged both when functions are idle or active). The standard charges that apply to on-demand instances are additionally applied for actively running function executions, albeit contextually \textbf{applied to the ``always-ready''} instance cost. We let \(v\) be the total baseline cost, with \(r_{total}\) being the total amount of memory provisioned for all the selected instances, \(T\) the total number of seconds the instances are provisioned for and \(p_b\) the instance baseline GB-second price. Note that \(AR\) refers to the always-ready instance cost: 

\begin{equation}
\label{eqn:flexalwayson}
\begin{aligned}
v &= \frac{r_{total}}{1024} \cdot T \cdot p_b \\[1ex]
C_{\text{AR}} &=
v
+
C_{\text{FLEX}}^{AR}(r,d)
\end{aligned}
\end{equation}

If functions have to use on-demand instances for scalability beyond already provisioned instances then the total cost would be \(C_{AR}\) added to \(C_{FLEX}\). Microsoft's final plan using provisioned concurrency is the Premium Plan \cite{azurefunctionspremiumplan}. With this approach, the user has permanently ``always-ready'' instances that are ready to serve requests. The user pays for the time the CPU core seconds and the time the memory on the dedicated resources are running, with no invocation fee charged. This cost is paid across all their chosen instances, resulting in simpler billing, but also means that the end-user is paying for instance(s) to be always running, regardless of whether they're processing requests or not (similar to a traditional IaaS VM). The Premium Plan must have a minimum of one instance enabled. The equation for the premium plan is therefore as follows: 

\begin{equation}
\label{eqn:azure_premium}
C_{\text{premium}}
=
\sum_{j=1}^{J}
\left(
r_j \cdot T \cdot p_{mem}
+
c_j \cdot T \cdot p_{cpu}
\right)
\end{equation}

where \(j\) is the specific instance, \(r_{j}\) is the configured memory amount, \(c_{j}\) denotes the number of CPU cores for the instance, \(T\) is the time the instance was in use and \(p_{m}\) and \(p_{cpu}\) are the price of the memory and CPU cores, respectively.

\subsubsection{AWS}

With AWS's Provisioned Concurrency, a user enables this feature on their existing function(s) \cite{awslambdapricing, awsprovisionedconcurrency}. Note that provisioned concurrency attracts a different pricing tier than standard Lambda execution prices. Similar to Azure, the user pays for the period of time that provisioned concurrency is activated, the function execution duration, in addition to an invocation fee, all at provisioned concurrency prices. If usage were to exceed provisioned concurrency resources, then standard on-demand Lambda instances are used. Note that whilst duration of function execution is calculated to the nearest millisecond \(t(d)\), the duration of a provisioned concurrency instance running (which we'll call \(T_{b}\)) is instead rounded up to the nearest five minutes. We can express the provisioned concurrency equation therefore as follows:

\begin{equation}
\label{eqn:awspc}
\begin{aligned}
T_b 
&= \left\lceil \frac{T}{300} \right\rceil \cdot 300 \\[1ex]
v_{\text{AWSPC}} 
&= \underbrace{k \cdot \frac{r}{1024} \cdot T_b \cdot p_{pc}}_{\text{Provisioned Concurrency}}\\[1ex]
C_{\text{AWSPC}} &=
v_{\text{AWSPC}}
+
C^{PC}_{\text{DEF}}(r,d)
\end{aligned}
\end{equation}

where \(k\) is the number of provisioned concurrency instances, \(p_{pc}\) is the cost per GB-second of running the instance and \(T_{b}\) is the duration for running the instance with \(T\) being given in seconds. The standard Equation \ref{eqn:faaspricing} is used to calculate the execution and invocation cost of running functions, contextually applied to using Provisioned Concurrency \(PC\) costs.

\subsubsection{Google}

Google's equivalent of pre-provisioned functions is called \textit{min-instances} \cite{googlecloudrun_min_instances}. For \textit{request-based} billing, Google charges for both CPU and Memory albeit at a lower \textit{idle} cost when not used \cite{googlecloudrun_min_instances}. When the instance is being used, standard billing \(C^{RB}_{GCF}\) is applied. If the user were to use \textit{instance-based} billing instead (with min-instances), they are charged the full rate for the instance regardless of whether it is actively processing requests or not. The instance-based billing for this mode is exactly the same as \(C^{IB}_{\text{GCF}}\) in Equation \ref{eqn:faaspricing_google_combined} multiplied by the number of instances \(k\) the user provisions. We model the min instances mode for Google Cloud Run Functions as follows:

\begin{equation}
\label{eqn:gcf_min_full_cidle}
\begin{aligned}
C^{RB}_{\text{GCFPC}}
&= k \cdot T \cdot \Bigg(
      \frac{r}{1024} \cdot \, p^{RB}_{m_{\mathrm{idle}}}
      + c \cdot \, p^{RB}_{cpu_{\mathrm{idle}}}
    \Bigg)    
\end{aligned}
\end{equation}

where \(p^{RB}_{cpu_{idle}}\) and \(p^{RB}_{m_{idle}}\) are the costs for idle usage of the provisioned resources for \textit{request-based} billing. When actively processing requests, we switch to using \(C^{RB}_{\text{GCF}}\) and then switch back to \(C^{RB}_{\text{GCFPC}}\) when idle again. Note that instance-based billing is not supported by the 1st generation environment of Google Cloud Functions.

\subsection{IaaS backed FaaS}

For IaaS backed FaaS, AWS Lambda Managed Instances \cite{awslambdapricing} has a hybrid billing model combining elements of FaaS pricing and traditional IaaS pricing. AWS removes the duration based billing for function execution/resource usage, but keeps the invocation fee. It then charges a compute management fee premium, which is 15\% applied on top of the price of the EC2 instances chosen. This is in addition to the standard per hour charge to use the instance in the first place. We model the managed instance pricing equation \(C_{AWSMI}\) as follows:

\begin{equation}
\label{eqn:awsmi_final}
C_{\text{AWSMI}}
=
\underbrace{
k \cdot T \cdot 1.15 \cdot p_{\text{EC2}}
}_{\text{Managed instance cost}}
+
\underbrace{
n \cdot p_i
}_{\text{Invocation}}
\end{equation}

where 1.15 represents the 15\% cost added to the price of the EC2 instance \(p_{EC2}\) and \(n\) are the number of invocations. For Azure's IaaS backed Faas \cite{azure_functions_dedicated_plan}, they make use of App Service Plans (ASPs) \cite{azure_app_service_hosting_plans}, Azure's primary use of billing for their VM infrastructure. The user selects the number of resources they require (VM type, VM resources, location, etc.) and then assign applications to the plan \cite{azure_app_service_hosting_plans}. Note that the user has a choice of using dedicated VMs for their ASPs (billed at standard IaaS rates) or shared infrastructure (where the user's app receives a quota of CPU minutes on shared infrastructure).\\

\subsection{Durable Functions}

Durable Functions pricing for AWS is more involved as it comprises several components. As mentioned in Section \ref{complexscenarios}, users can build complex applications using primitives such as \textit{waits} or \textit{retries} as well as use storage for storing intermediate operations, etc. Multiple functions can also be combined/chained into a workflow \cite{aws_lambda_durable_functions}. AWS splits its billing into four components. Users are billed for standard function invocation any time a function is invoked and run like in Equation \ref{eqn:faaspricing}. Usage of what AWS describes as durable operations, such as steps, callbacks, waits, etc. is then additionally billed per million operations. Finally, for data written during operations, this is charged per GB with data written retained during and after execution charged in a prorated fashion per GB-month. We can therefore express the pricing model equation for Durable Functions on AWS as follows:

\begin{equation}
\label{eqn:awsdurable_storage_aws}
\begin{aligned}
C_{\text{AWSDurable}}
&= 
C_{DEF}(r,d) +
\underbrace{N_{\text{ops}} \cdot p_{\text{ops}}}_{\text{Durable operations}} \\
&\quad + 
\underbrace{D_{\text{written}} \cdot p_{\text{write}}}_{\text{Data written during execution}} \\
&\quad + 
\underbrace{D_{\text{retained}} \cdot p_{\text{retention}}}_{\text{Data retained after completion}}
\end{aligned}
\end{equation}

where \(N_{ops}\) and \(p_{ops}\) refers to the number of durable operations utilized and the cost of them, whilst data written and retained (\(D_{written}\) and \(D_{retained}\)) also attract their own pricing points. Azure Durable Functions are charged in a very different way to AWS \cite{azure_durable_billing}. If the user is using the Microsoft Consumption Plan (due to expire in 2028) then they are charged an additional fee for function replays (charged as a standard function invocation). However if they are using any of the other plans, no extra charge is levied for using Durable Functions. The only other element that is charged regardless of plan is for use of Azure Storage for state/transaction persistence. This isn't an extra cost, per se, of function execution as Azure Storage costs are global to the user's account and as such are charged as part of the user's standard Azure subscription \cite{azure_durable_billing}.\\ 

We have now modelled the pricing equations for the various pricing strategies employed by the FaaS providers. Next, using these equations, we will focus on charting and comparing the estimated cost of running workloads across the three providers.

\begin{figure*}[!htbp]
\centering
\caption{Function Execution Cost (All Providers) - Standard and Provisioned}
\label{fig:estimatedcosts}

\begin{subfigure}{0.3\textwidth}
    \centering
    \includegraphics[width=\linewidth]{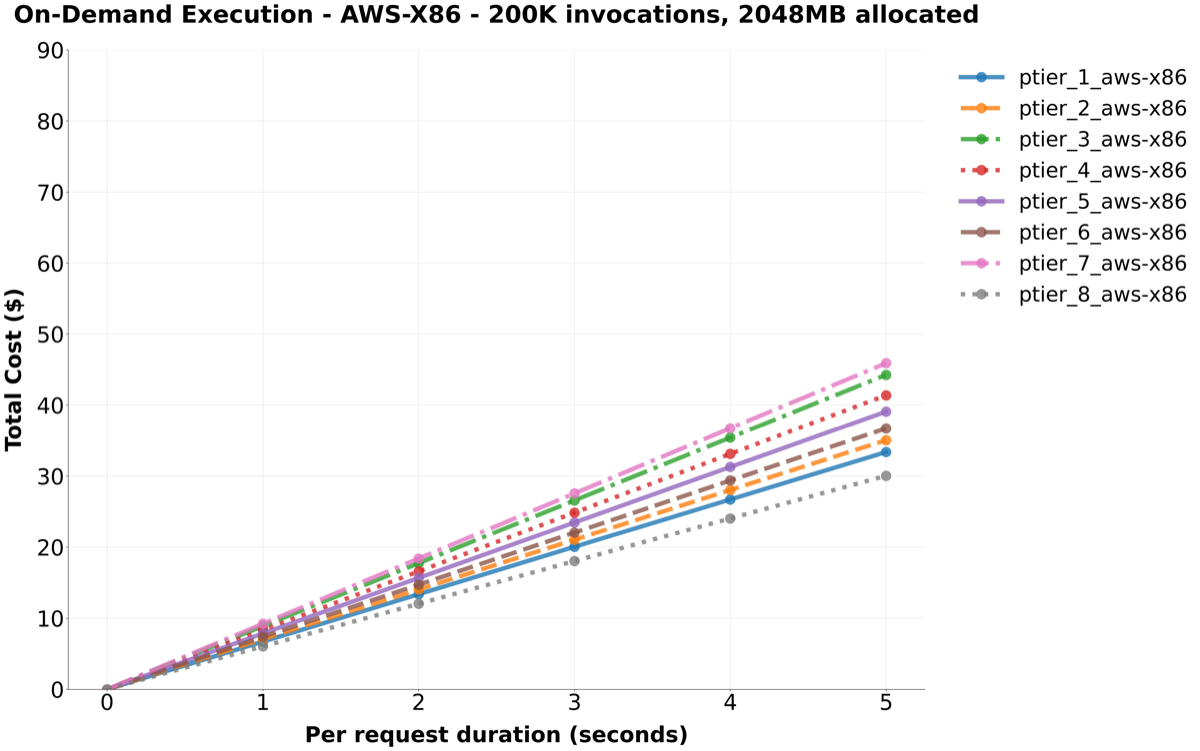}
    \caption{AWS Lambda (x86) - Standard}
    \label{fig:awsx86standard}
\end{subfigure}\hfill
\begin{subfigure}{0.3\textwidth}
    \centering
    \includegraphics[width=\linewidth]{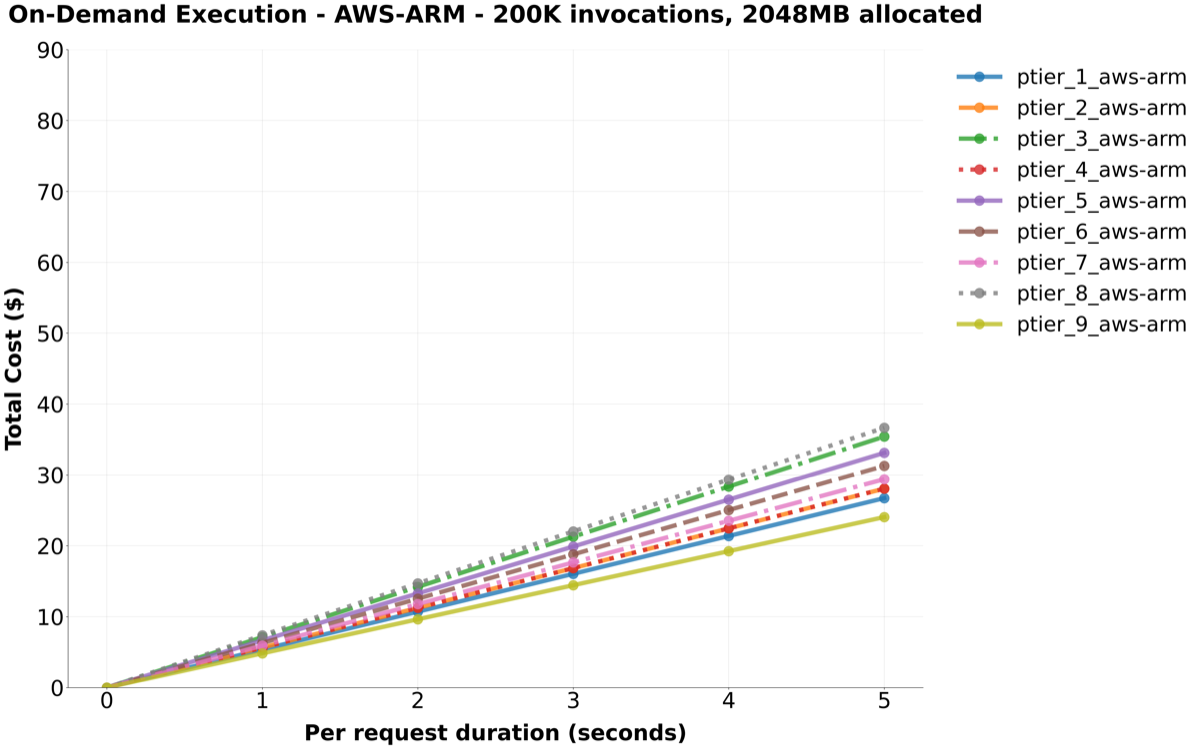}
    \caption{AWS Lambda (ARM) - Standard}
    \label{fig:awsarmstandard}
\end{subfigure}\hfill
\begin{subfigure}{0.3\textwidth}
    \centering
    \includegraphics[width=\linewidth]{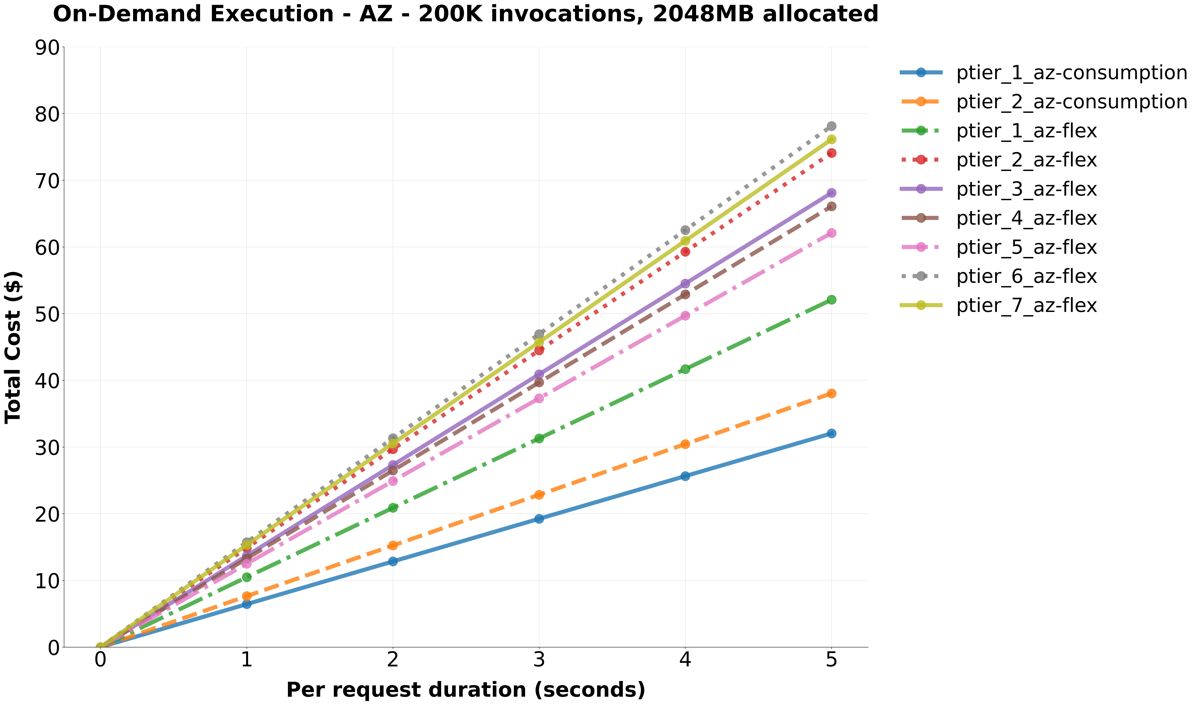}
    \caption{Azure - Standard}
    \label{fig:azstandardpricing}
\end{subfigure}

\vspace{0.4cm}

\begin{subfigure}{0.3\textwidth}
    \centering
    \includegraphics[width=\linewidth]{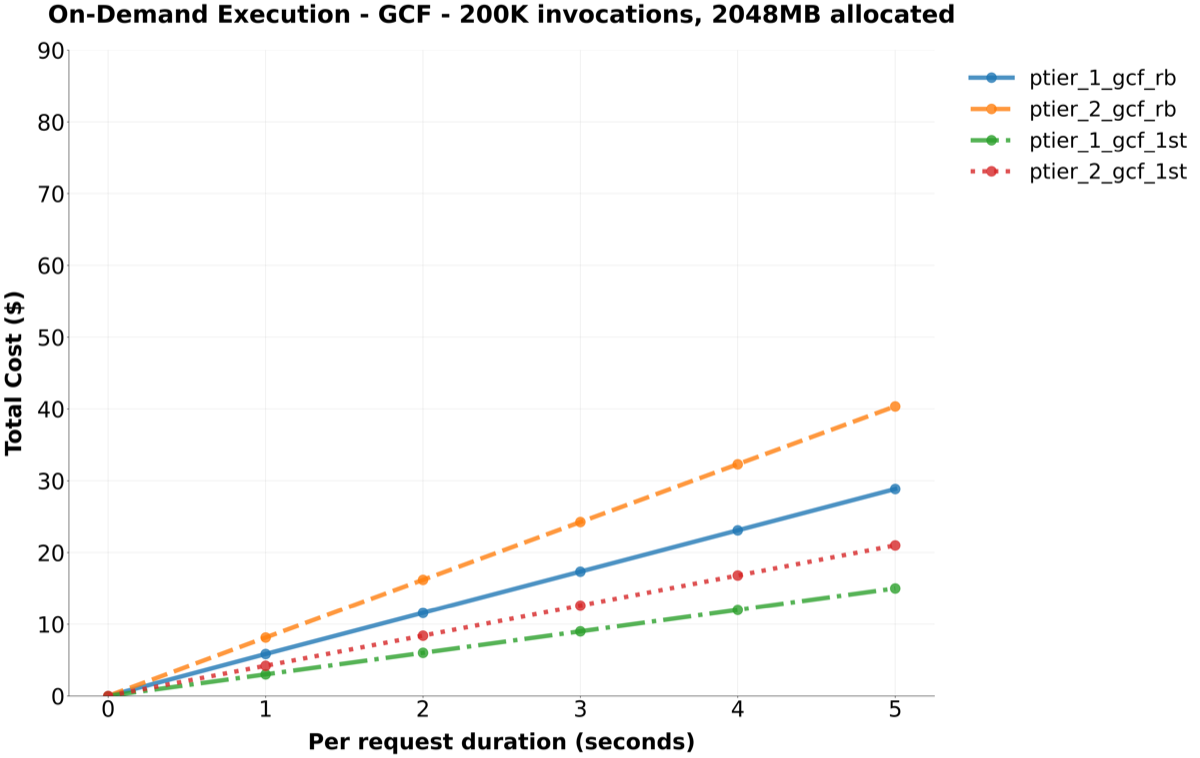}
    \caption{GCF - Standard}
    \label{fig:gcfstandard}
\end{subfigure}\hfill
\begin{subfigure}{0.3\textwidth}
    \centering
    \includegraphics[width=\linewidth]{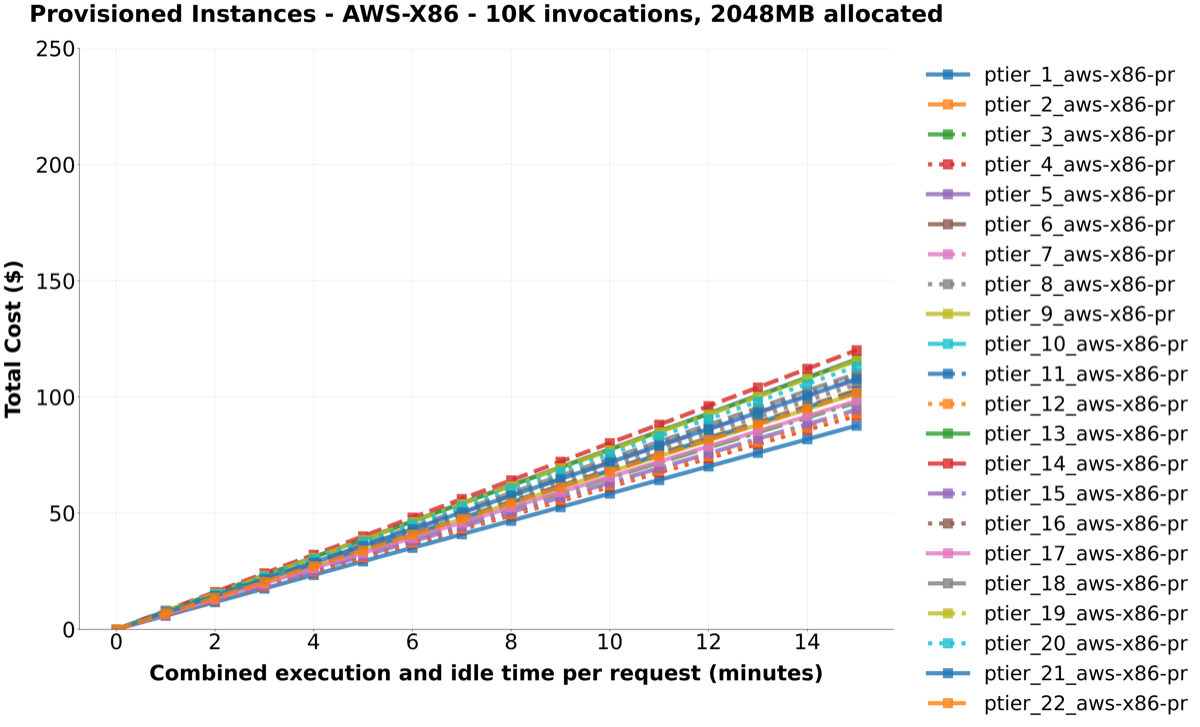}
    \caption{AWS (x86) - Provisioned}
    \label{fig:awsx86prstandard}
\end{subfigure}\hfill
\begin{subfigure}{0.3\textwidth}
    \centering
    \includegraphics[width=\linewidth]{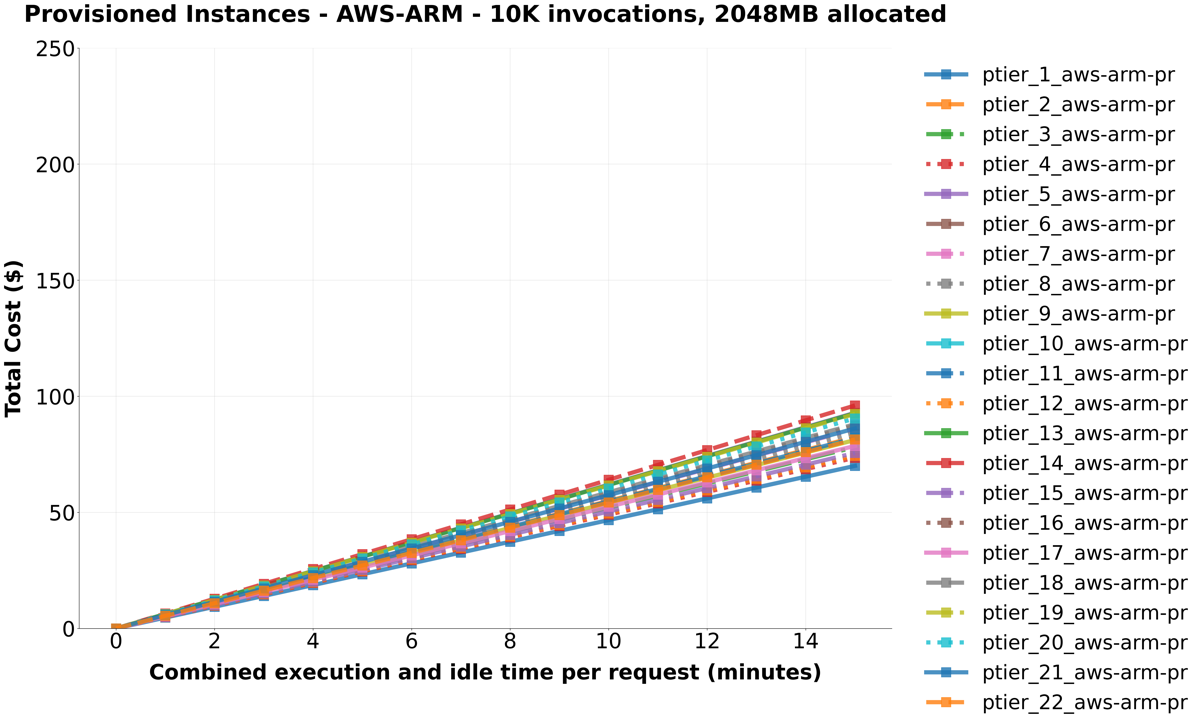}
    \caption{AWS (ARM) - Provisioned}
    \label{fig:awsarmprstandard}
\end{subfigure}

\vspace{0.4cm}

\begin{subfigure}{0.3\textwidth}
    \centering
    \includegraphics[width=\linewidth]{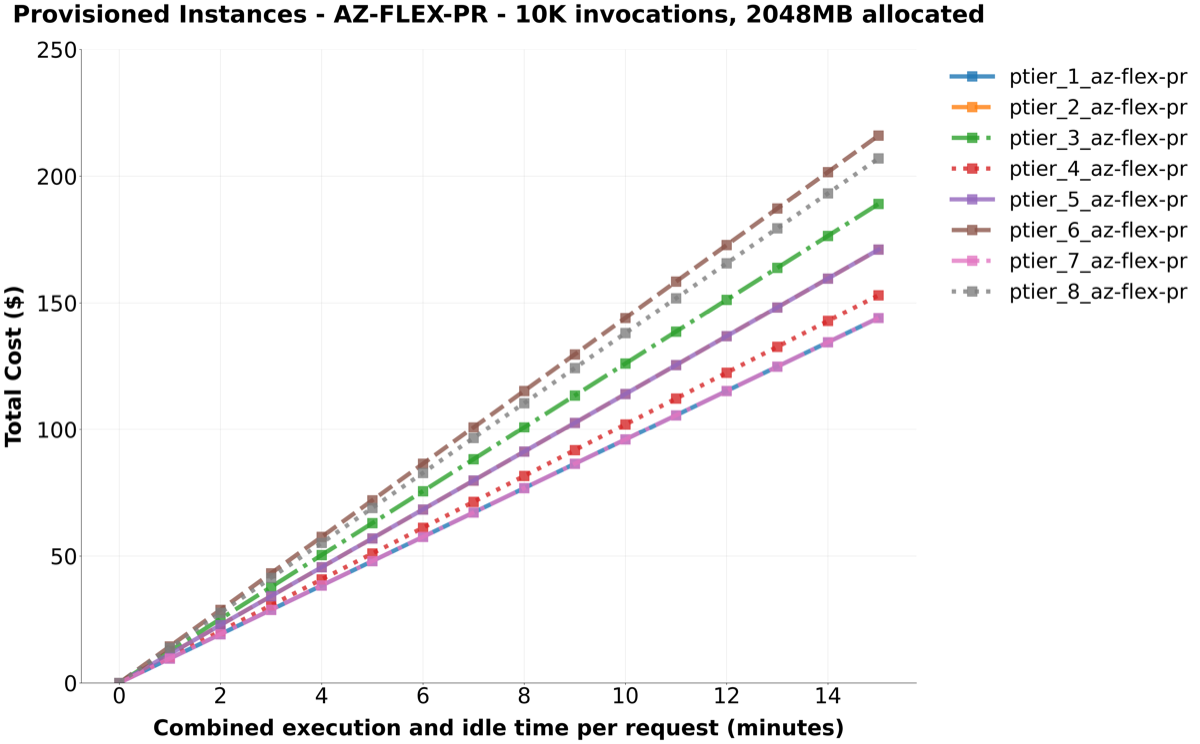}
    \caption{Azure Flex - Provisioned}
    \label{fig:azflexprstandard}
\end{subfigure}\hfill
\begin{subfigure}{0.3\textwidth}
    \centering
    \includegraphics[width=\linewidth]{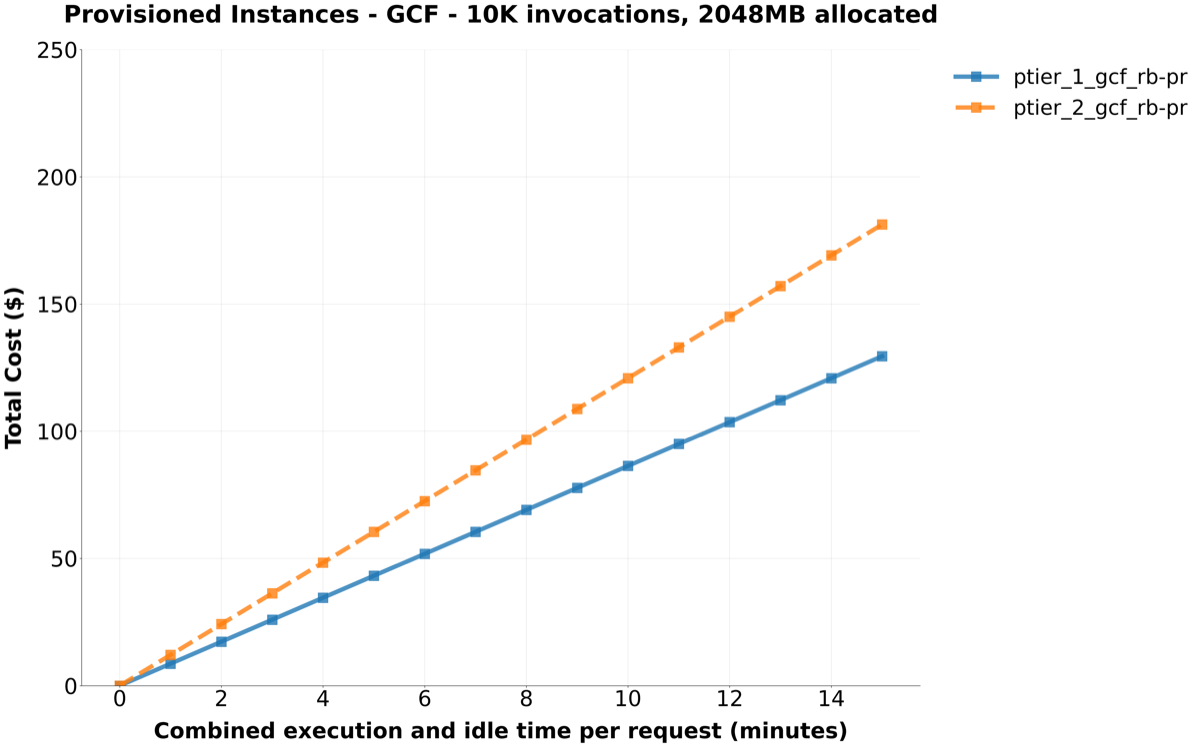}
    \label{fig:gcfpr}
    \caption{GCF - Provisioned}
\end{subfigure}

\end{figure*}

\section{Analysis} \label{analysis}

In our analysis, we will focus on comparing the estimated cost of running functions on both standard on-demand FaaS as well as using pre-provisioned instances, as these are both features supported by \textit{all} three providers. All three providers support multiple regions for deployment \cite{ekwe2024state}, which attracts different pricing tiers depending on the region selected. As such, we have grouped and classified the various regions into common pricing tiers for each provider. For example, \textit{$ptier{_2}_{aws-x86}$} refers to a price point of \(\$2.10 \times 10^{-7}\) for the function invocation fee and \(\$1.75 \times 10^{-5}\) for the memory compute cost to the Mexico, Israel and New Zealand regions \textit{mx-central-1, il-central-1 and ap-southeast-6} \cite{awslambdapricing}. In contrast, \textit{$ptier{_3}_{aws-x86}$} applies a cost of \(\$2.70 \times 10^{-7}\) and \(\$2.21 \times 10^{-5}\) to the invocation and compute cost respectively for functions run in the South Africa \textit{af-south-1} region \cite{awslambdapricing}. Some providers, such as AWS and Azure (specifically on their Flex Consumption Plans) have multiple pricing tiers/region classifications, whereas providers such as Google and Azure (Consumption Plan Standard) have only two pricing tiers applied across supported regions. For our comparison of providers using standard on-demand pricing, the scenario we use is a user running a function workload that is invoked 200K times with 2048MB of memory allocated to the function. For ease of comparison, we assume that all the function invocations have uniform running time and all invocations use all 2048MB of memory. Therefore, on the x-axis and y-axis of our graphs, (Figure \ref{fig:estimatedcosts}), the x-axis denotes all 200K functions running for n seconds with the y-axis showing the total cost of these.

\subsection{Standard On-Demand Pricing}

As we can see in Figure \ref{fig:estimatedcosts}, the first thing we notice is the significant number of regional price tiers for both AWS and Azure with only 2 for GCF. For AWS, there are 8 different pricing tiers applied to all 33 regions that support x86 FaaS deployments, with 9 different pricing tiers for ARM FaaS deployments (Figures \ref{fig:awsx86standard} and \ref{fig:awsarmprstandard}). The majority of AWS regions (21) are charged at \(\$2.00 \times 10^{-7}\) and \(\$1.67 \times 10^{-5}\) for the invocation and memory costs respectively (\textit{$ptier{_1}_{aws-x86}$}). This price point is applied to the entire of the United States and Canada, with the majority of regions from Europe and a significant number for Asia Pacific included. The \textit{$ptier{_1}_{aws-x86}$} represents the second cheapest price tier for running FaaS functions. The cheapest, \textit{$ptier{_8}_{aws-x86}$}, is applied to the Malaysia, Taiwan and Thailand regions at a price point of \(\$1.80 \times 10^{-7}\) and \(\$1.50 \times 10^{-5}\) for the invocation and memory costs respectively. The most expensive pricing tier (\textit{$ptier{_7}_{aws-x86}$}) is applied to the Hong Kong region at a price point of \(\$2.80 \times 10^{-7}\) for \(p_{inv}\) and \(\$2.29 \times 10^{-5}\) for \(p_{mem}\) respectively. The nine remaining regions are spread out amongst the 5 remaining price tiers, with a mode value of 1 region assigned per pricing tier. Switching to the ARM Lambda deployments (Figure \ref{fig:awsarmstandard}, we see cheaper costs for running functions across the board. The invocation \(p_{inv}\) costs remain constant regardless of ARM or x86 deployment, however there are significantly cheaper \(p_{mem}\) costs for running workloads. Examining the average total cost of running x86 and ARM deployments across all regions reveals ARM deployments being on average 20.52\% cheaper to run than x86. The region price tiers are exactly the same across the ARM and x86 deployments. The only point to note is that the \textit{$ptier{_2}_{aws-x86}$} (consisting of the Mexico, Israel and New Zealand regions) have the same invocation price point for x86 but for ARM there's a slight increase for Israel and New Zealand deployments vs Mexico, putting Israel and New Zealand into their own separate price tier. The cheapest regions for deploying ARM functions are the Malaysia, Thailand and Taiwan regions (\(\$1.80 \times 10^{-7}\) for \(p_{inv}\) and \(\$1.20 \times 10^{-5}\) for \(p_{mem}\) respectively) with the most expensive being the Hong Kong (\(\$2.80 \times 10^{-7}\) for \(p_{inv}\) and \(\$1.83 \times 10^{-5}\) for \(p_{mem}\)). \\

Moving to Microsoft Azure, we see an interesting set of results (Figure \ref{fig:azstandardpricing}). First, for the legacy Microsoft Consumption Plan, we only see two pricing tiers across the 47 regions supported by Azure. The majority of regions (45) are included in the \textit{$ptier{_1}_{az-consumption}$} price tier at a price point of  \(\$2.00 \times 10^{-7}\) for \(p_{inv}\) and \(\$1.60 \times 10^{-5}\) for \(p_{mem}\). Two regions \textit{malaysia-west} and \textit{indonesia-central} attract a slightly higher \(p_{mem}\) cost of \(\$1.90 \times 10^{-5}\). For the newer Flex Consumption Plan however, Microsoft follows more of the AWS model, having a wide range of pricing tiers for all of its regions - in Azure's case, 7 tiers. With this increase in tiers, also comes a \textit{significant} increase in pricing (nearly 2x more expensive than Consumption), with the cheapest pricing tier (\textit{$ptier{_1}_{az-flex}$}) charging  \(\$4.00 \times 10^{-7}\) for \(p_{inv}\) and \(\$2.60 \times 10^{-5}\) for \(p_{mem}\) with the most expensive price tier (\textit{$ptier{_6}_{az-flex}$}) having a price point of \(\$6.00 \times 10^{-7}\) for \(p_{inv}\) and \(\$3.90 \times 10^{-5}\) for \(p_{mem}\). The majority of regions are in the \textit{$ptier{_1}_{az-flex}$} and \textit{$ptier{_2}_{az-flex}$} pricing tiers (37), with the remaining 8 spread across the remaining 5 pricing tiers. Note also, that for Flex Consumption, only 45 regions are supported as opposed to 47 for the legacy Consumption Plan.\\

For Google Cloud Functions (Figure \ref{fig:gcfstandard}), for both 1st gen legacy and current pricing, two pricing tiers are applied across all supported regions. For the 1st generation GCF, only 21 total regions are supported to deploy functions whilst with the current generation Cloud Run Functions, 42 regions are supported \cite{googlecloudfunctionspricing}. As mentioned earlier in Section \ref{pricing-models}, Google Cloud offers three different price points for the invocation, memory and CPU. The \textit{$ptier{_1}_{gcf-1st}$} has a price point of \(\$2.50 \times 10^{-6}\) for \(p_{mem}\) and \(\$1.00 \times 10^{-5}\) for \(p_{cpu}\) with an increase to \(\$3.50 \times 10^{-6}\) for \(p_{mem}\) and \(\$1.40 \times 10^{-5}\) for \(p_{cpu}\) for the second pricing tier for the 1st generation pricing. With the \textit{$ptier{_1}_{gcf-rb}$} and \textit{$ptier{_2}_{gcf-rb}$} current generation pricing, we see an increase in running costs. In the first pricing tier, the \(p_{mem}\) remains the same at \(\$2.50 \times 10^{-6}\) but increases significantly to \(\$2.40 \times 10^{-5}\) for \(p_{cpu}\). In the second pricing tier, both the \(p_{cpu}\) and \(p_{mem}\) increases even more to \(\$3.50 \times 10^{-6}\) for \(p_{mem}\) and \(\$3.36 \times 10^{-5}\) for \(p_{cpu}\) respectively. Across both price generations, the invocation cost per request \(p_{inv}\) is \(\$4.00 \times 10^{-7}\). Regions are more evenly spread out in 1st generation and 2nd generation pricing across the two pricing tiers with regions from all continents included in both tiers. For the 1st generation pricing there are a total of 21 regions supported - 9 regions included in the first pricing tier vs 12 in the second. For the current, second generation pricing, this jumps to 23 in tier 1 and 19 in tier 2. \\

Comparing the costs of running functions across all the providers using standard on-demand billing, we note the following. The cheapest platform to run workloads on is the 
Google Cloud Functions (1st gen) option followed by the AWS ARM option, with the newer Microsoft Azure Flex Consumption Plan being the most expensive. The Flex Consumption Plan is 1.78x more expensive than AWS x86, 2.25x more expensive than running on ARM, 1.97x more expensive than Google Cloud Functions (current generation) and 3.78x more expensive than Google Cloud Function's 1st generation pricing. When compared against Microsoft's own legacy Consumption Plan, the Flex plan is 1.94x more expensive. AWS ARM, the second cheapest provider, is 1.2x cheaper than AWS's x86 platform and 1.14x cheaper than GCF's second generation pricing. Microsoft's Consumption Plan matches the current generation Google Cloud Function's pricing, and is also 1.1x cheaper than AWS's x86 pricing. However, when compared to AWS's ARM platform, it becomes 1.16x more expensive.

\subsection{Pre-Provisioned Instances} \label{pre-provisioned-instances-results}

For Pre-Provisioned Instances, we model a different scenario for our comparison. As we mentioned in our exploration of the Pre-Provisioned pricing models (Section \ref{pre-provisioned-functions-models}), different rates are charged when instances are idle and not processing requests, vs when they are active. As such we model a combined scenario of active and idle time to show the differences in providers. We model: a user running 10K requests for their function with 2 pre-provisioned instances with each function execution running for the same time and using all resources available. In the standard on-demand graphs, each time period denoted all 200K functions each running for the time period shown - for example 1 second on the x-axis mapped 200K functions running for 1s each with 2048MB of memory. In the case of our pre-provisioned instances however, each time period on the graphs (Figures \ref{fig:awsx86prstandard} - 1h) illustrates a 50-50 split between actual running time and idle time. That is, to give an example, 10m on the x-axis illustrates 10K functions running for 5 minutes each in addition to 5 minutes of idle time per request. As with the standard on-demand experiment 2048MB is allocated to the function in our scenario.\\

The first thing we notice for AWS's results (Figure \ref{fig:awsx86prstandard} and Figure \ref{fig:awsarmprstandard}) are the significant increase in pricing tiers for Provisioned Concurrency - 22 pricing tiers vs the 7/8 we saw in the standard On-Demand pricing. There are also fewer regions per pricing tier, with the modal value for a tier being just one region. By means of comparison, if we were to examine the individual costs of \(p_{inv}\), \(p_{mem}\) and \(p_{pc}\) at AWS provisioned concurrency rates, we see the cheapest tier is the \textit{$ptier{_1}_{aws-x86-pr}$} consisting North Virginia, Ohio and Oregon with \(p_{inv}\) being \(\$2.00 \times 10^{-7}\), 
\(p_{mem}\) being \(\$9.7222 \times 10^{-6}\) and \(p_{pc}\) being
\(\$4.1667 \times 10^{-6}\). The most expensive pricing tier \textit{$ptier{_{14}}_{aws-x86-pr}$} consisting of the Hong Kong region is priced at  
\(\$2.80 \times 10^{-7}\), \(\$1.33358 \times 10^{-5}\) and \(\$5.7153 \times 10^{-6}\) for \(p_{inv}\), \(p_{mem}\) and \(p_{pc}\) respectively. Switching to ARM, just as with On-Demand Execution, we notice significantly cheaper costs across the board. The cheapest/most expensive price tiers remain the same albeit with significantly reduced costs. For tier 1, \(p_{inv}\) becomes \(\$2.00 \times 10^{-7}\), 
\(p_{mem}\) becomes \(\$7.7778 \times 10^{-6}\) and finally \(p_{pc}\) becomes
\(\$3.3334 \times 10^{-6}\). For tier 14, \(p_{inv}\) increases to \(\$2.80 \times 10^{-7}\), \(p_{mem}\) becomes \(\$1.06686 \times 10^{-5}\) and finally \(p_{pc}\) becomes
\(\$4.5722 \times 10^{-6}\).\\

Examining Microsoft's ``always-ready'' instances contained in Azure Flex Consumption (Figure \ref{fig:azflexprstandard}), we see a relatively small set of price tiers (8). However, and similar to the on-demand pricing, Azure represents a significant increase in price compared to AWS. The cheapest pricing tier \textit{$ptier{_1}_{az-flex-pr}$} has a baseline \(p_{b}\) of \(\$4.00 \times 10^{-6}\), a compute cost \(p_{mem}\) of \(\$1.60 \times 10^{-5}\) and an invocation fee \(p_{inv}\) of \(\$4.00 \times 10^{-7}\). This tier includes 14 regions spanning several continents. The most expensive region, \textit{south-india}  contained in \textit{$ptier{_6}_{az-flex-pr}$} marks a significant increase in all three of these costs at \(\$6.00 \times 10^{-6}\), \(\$2.40 \times 10^{-5}\) and \(\$6.00 \times 10^{-7}\) for the baseline, compute and invocation costs respectively. Whilst the ``always-ready'' instances represent an increase on AWS, the Premium Plan can become even more expensive. The nuance to note however with the Premium Plan is that it is similar to a traditional VM, with the cheapest montly cost being over \$110 for vCPU and memory GB costs being over \$8 \cite{microsoftazurefunctionspricing}. There are also only three instance sizes supported (3.5, 7 and 14GB of memory) and 1, 2 and 4 cores for this plan. There is however no invocation fee charged - users simply pay for the time the instance is provisioned.\\

Finally, with Google Cloud Run Functions, we see similar patterns to what we saw for standard on-demand pricing. We see the same two pricing tiers, with the first pricing tier being significantly cheaper than the second. What is interesting to note is that the \textit{idle time} and \textit{active time} are in some cases the same in some price tiers. By way of reminder, we noted that for the first google price tier (\textit{$ptier{_1}_{gcf-rb}$}), the \(p_{inv}\) cost was \(\$4.0 \times 10^{-7}\), \(p_{mem}\) was \(\$2.5 \times 10^{-6}\) and finally \(p_{cpu}\) was \(\$2.4 \times 10^{-5}\). The idle time rate for CPU and Memory is exactly the same at \(\$2.5 \times 10^{-6}\). This is different in the second pricing tier, with the invocation fee being \(\$4.0 \times 10^{-7}\), the memory being \(\$3.5 \times 10^{-6}\) and finally the CPU cost being \(\$3.36 \times 10^{-5}\). The idle time for memory and CPU are both the same at \(\$3.5 \times 10^{-6}\). With the \textit{instance-based} billing, similar to Azure Premium, this is charged for the full lifetime of the instance rather than being on-demand. For tier 1, The memory cost is \(\$2.0 \times 10^{-6}\), CPU cost is \(\$1.8 \times 10^{-5}\) and (unlike Microsoft Azure Premium) an invocation fee is also charged at \(\$4.0 \times 10^{-7}\). For tier 2, the invocation fee remains the same but the memory cost is \(\$2.0 \times 10^{-6}\) and cpu cost is \(\$2.16 \times 10^{-5}\).
\\

Comparing between all three providers, as with on-demand pricing, AWS ARM represents the cheapest running cost for functions. AWS x86 is the second cheapest provider, with Google's \textit{request-based} billing being the third cheapest. The most expensive provider is Azure Flex. Azure Flex is 1.12x more than GCF Request-based billing, 1.67x more expensive than AWS x86 and 2.09x more than AWS ARM.

\section{Observations} \label{observations}

In our analysis, we explored pricing differences of running our two function scenarios 
across all the on-demand providers (both in standard and pre-provisioned modes). We make three key observations from our work:\\

\begin{mdframed}
\textit{Observation 1}: There are significant price differences across providers which the end-user must identify to ascertain the most cost-effective provider for their workload.
\end{mdframed}

Our analysis has shown the significant cost differences between the various providers for FaaS workloads. On the surface, Google's 1st gen pricing is the cheapest to deploy functions in. However, this is a legacy platform and is not generally supported for new accounts running new workloads. Excluding 1st gen GCF pricing, AWS represents the cheapest cost for running workloads, whilst Azure (specifically its new Flex Consumption Plan) is the most expensive. Examining AWS in more detail revealed that whilst AWS's ARM platform is consistently the cheapest, the legacy Azure Consumption Plan (still available for running new function workloads) and Google's 2nd gen Cloud Functions pricing represents the next two cheapest platforms to run on-demand workloads with. Microsoft Azure's new Flex Consumption Plan is significantly more expensive than any of the other providers, and once the older Consumption Plan is removed in 2028 \cite{azurefunctionsconsumptionplan}, will represent a significantly higher price point overall than the other two providers AWS and GCF. Switching to provisioned mode shows a different set of results. In this area, both AWS's ARM and x86 represent the cheapest cost for running function workloads, with Google and Azure being significantly more expensive (Azure Flex Consumption Plan). End-users must explore these costs further, and particularly in the era of multi-cloud, can exploit these different providers' offerings for overall cost reduction \cite{georgios2021exploring}.\\

\begin{mdframed}
\textit{Observation 2}: Regional differences are significant across FaaS offerings, creating new research opportunities for cost exploitation cross-region.
\end{mdframed}

For Google Cloud Functions (1st and 2nd gen), their function pricing was grouped into two tiers encompassing all their regions. However, for AWS, and particularly for Azure, we saw a significant number of pricing tiers and as such varying costs cross-region. In the case of AWS particularly on the provisioned concurrency mode, we noted 22 different pricing tiers applied to the various regions with, in a large number of cases, only 1 region attracting that particular price point. That was very different to the standard on-demand pricing where, we had significantly fewer pricing tiers and as such the majority of tiers included multiple regions within the tier with fewer tiers containing 1-3 regions per tier. This area requires further exploration and also opens up opportunities for price exploitation cross-region as seen in other areas of the cloud such as spot instances \cite{ekwe2018location}. There is a trade-off to this approach of course, as the lowest priced region may not always be the optimal region for certain workloads (for example requiring lowest latencies).\\

\begin{mdframed}
\textit{Observation 3}: For Azure and Google, the newer generation pricing plans are significantly more expensive than first generation pricing iterations.
\end{mdframed}

Google's newer second generation pricing and Azure's Flex Consumption Plan represented a nearly 2x increase on previous generations. Part of this was due to a change in billing model (for example with Azure's minimum billing of 1s) as well as increases in overall pricing of individual components (cpu, mem, etc.). Studies examining a price-performance ratio of these various providers would help in understanding this area more.

\section{Related Work} \label{relatedwork}

There have been a number of works that have sought to explore and understand the differences and strategies in serverless pricing. The work \cite{liu2023demystifying} sought to understand serverless pricing strategies and compared serverless to IaaS approaches. Their work observed that serverless was not always the cheapest when compared to IaaS for certain workloads, and then proposed a new model to avoid expense explosion for users. Other works \cite{ghorbian2026survey, lin2026demystifying, perez2018cost, hamza2023understanding} have also sought to analyse and understand the key models and differences in pricing between the providers. As a means of solving some of the issues in pricing for serverless, there has been the creation and implementation of new optimization and predictive frameworks for serverless performance/pricing \cite{jarachanthan2022astrea, elgamal2018costless, lin2020modeling, eismann2020predicting}. From an industry point of view, we've seen blogs from companies such as cloud security platform Wiz \cite{wiz_azure_vs_aws_cost_2025}, Intercept \cite{kroeze_azure_vs_aws_pricing_2025} on Azure vs AWS costs, in addition to individual developers \cite{inigo_sanchez_aws_lambda_vs_gcf_pricing_2024, hanhaliuk_gcf_vs_lambda_2024}.

\section{Conclusion} \label{conclusion}

In this paper, we have introduced \textit{Priceless}, giving the current state of the art in features offered by serverless providers (such as provisioned concurrency, IaaS backed FaaS, Durable Functions, etc.), modelling their various pricing equations and strategies, and finally proceeding to perform a cost comparison using two different workload scenarios. We explored the differences between running workloads on standard on-demand FaaS as well as using pre-provisioned instances. We also in a new addition to the state of the art noted that there are \textit{intra-regional} cost differences within the providers that can be exploited for cheaper function execution. Our work aims to provide the research and industry communities with a snapshot of estimated costs of running workloads on FaaS with future work/directions available in various sectors such as inter-region price exploitation or price-performance ratio analysis using real-world workloads on all the providers, amongst others.



\bibliographystyle{IEEEtran}
\bibliography{references.bib}

\end{document}